\documentclass[twocolumn,twoside,slac_two]{revtex4}
\usepackage{graphicx}
\usepackage{fancyhdr}
\usepackage{textcomp}
\begin{document}

\title{Evidence for VHE emission from SNR G22.7-0.2 region with H.E.S.S.}

\author{H. Laffon, B. Kh\'elifi, F. Brun}
\affiliation{Laboratoire Leprince-Ringuet,  Palaiseau , France}

\author{F. Acero, J. M\'ehault}
\affiliation{Laboratoire Univers et Particules, Montpellier , France}

\author{G. P\"uhlhofer}
\affiliation{Institut f\"ur Astronomie und Astrophysik, Universit\"at T\"ubingen, Germany}

\author{P. Eger}
\affiliation{Physikalisches Institut, Universit\"at Erlangen-Nuremberg, Germany}

\author{M. Jamrozy}
\affiliation{Astronomical Observatory, Jagiellonian University, Cracow, Poland}

\author{A. Djannati-Ata\"i}
\affiliation{Astroparticules et Cosmologie, Universit\'e Paris 7, France}

\author{A.-C. Clapson}
\affiliation{Max-Planck-Institut f\"ur Kernphysik, Heidelberg, Germany, now at EMBL Heidelberg, Germany}

\author{for the H.E.S.S. collaboration}
\begin{abstract}
Observations of the supernova remnant G22.7-0.2 in the field of view of W41 with the H.E.S.S. telescope array have 
resulted in evidence for a very high energy (VHE) gamma-ray emission from the direction of the object. The emission 
region is spatially coincident with molecular clouds visible in $^{13}$CO data, suggesting a hadronic origin of the 
TeV emission, although other scenarios brought by X-ray observations with the XMM-Newton satellite are also discussed. 
The latest results obtained on this SNR with H.E.S.S. and associated multi-wavelength information are presented here.
\end{abstract}

\maketitle

\thispagestyle{fancy}

\section{Introduction}

The supernova remnant (SNR) G22.7-0.2 shows a non-thermal ring in radio \cite{magpis} within a complex region and 
partially overlaps the remnant W41. 
G22.7-0.2 has an intrisic diameter of 26' in radio and neither its distance nor its age have been estimated. 
However, Spitzer observations \cite{spitzer} show enhanced infrared emission between G22.7-0.2 and W41, which might be 
interpreted as an interaction between the two remnants. An interstellar cloud surrounded and shocked by both remnants 
seems likely to produce the observed IR emission. In such an assumption, the distance of G22.7-0.2 should be close to the 
estimated distance of W41 of 4.2 kpc.

Observations carried out with the H.E.S.S. telescope array have led to an indication of a VHE gamma-ray excess in 
positional coincidence with a fraction of SNR G22.7-0.2 radio shell. 
The features of this  VHE emission are described in the second section. 

Non-thermal emission is found at the gamma-ray excess position by the Effelsberg telescope at 11 cm \cite{effels2695} 
and 21 cm \cite{effels1408} wavelengths. 
Additional multi-wavelength counterparts are found at this position, in particular in X-rays and in radio at the 
$^{13}$CO (J=1$\rightarrow$0) transition line energy. More details on these counterparts are given in the third section
 of this proceeding. The different considered scenarios that could explain the gamma-ray emission are discussed in the 
last section.
\vspace{0.35cm}

\section{H.E.S.S. observations}\label{hessobs}

\begin{figure}[htbp]
\includegraphics[width=85mm]{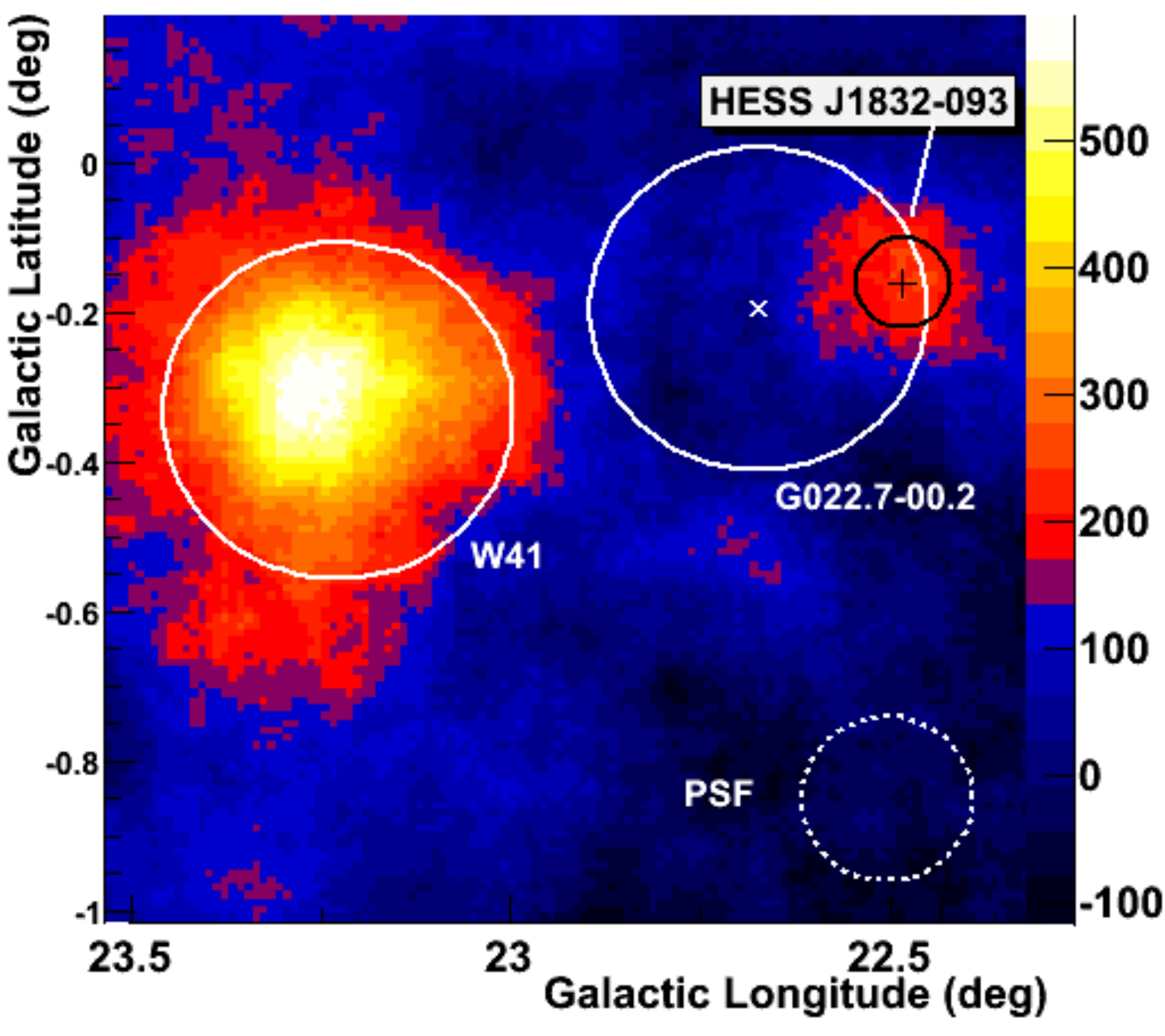}
\centering
\caption{H.E.S.S. correlated excess map to  the r$_{68}$ value of 0.11$^\circ$ at this position (represented by the 
dashed circle). The gamma emission from W41 is seen on the left of the image and the hot spot HESS J1832-093 on the 
right. Its best fitted position with corresponding errors is represented by the black cross and its intrisic size by a 
black circle. The apparent sizes of the SNR observed in radio are symbolized by white circles.}
\label{HESSexcess}
\end{figure}

Situated near the galactic plane in a field of view rich in gamma-ray sources, evidence for a new VHE gamma-ray source 
was found thanks to observations dedicated to study the neighbouring sources, such as W41 which is bright in 
VHE gamma-rays as well \cite{w41}.
To investigate such excess a detailed analysis of the region is performed.
A standard run selection procedure is used to remove bad quality observations. This results in a data set 
comprising 50 hours live time of observations taken from 2004 to 2010. Then, Hillas analysis with standard 
Hillas-based cuts \cite{crab} including a minimum charge of 80 photo-electrons in the shower images is applied to the 
data. The background is estimated with the ring-backgroung model, as described in \cite{crab}. 
The image shows evidence at a significance level of 5.8$\sigma$ pre-trials of a new compact VHE source dubbed 
HESS J1832-093. 
The corresponding excess map of this field of view is shown in Fig. \ref{HESSexcess}.

The average angular resolution r$_{68}$, defined as 68$\%$ containment of the point spread function (PSF), 
is 0.11$^\circ$ for these standard cuts at the source position and for these observations.
A two-dimensional Gaussian function is used to determine the position and size of the TeV emission. The best fitted 
position is $l= 22.48^\circ \pm 0.02_{\rm stat}^\circ$, $b= -0.16^\circ \pm 0.02_{\rm stat}^\circ$ in galactic coordinates, 
with a systematical pointing error of 0.01$^\circ$. 
The source appears slightly extended with an intrisic size of $0.06^\circ \pm 0.02^\circ$ estimated by using the 2D 
Gaussian function adjusted on the excess map and taking into account the PSF value at this~position. 

To produce the energy spectrum, only high quality data are used. Therefore the run selection is restricted to 
observations runs 
with four triggering telescopes only and with a pointing position less than 2.1$^\circ$ away from the source center. 
This results in a data set of 36 hours live time. The background is estimated with the reflected background 
model \cite{crab}, better suited for spectral studies.
Since the source is slightly extended, the region used to estimate the flux has to be large enough to contain all the 
signal from HESS J1832-093 but not too big to avoid signal contamination from W41. We thus choose a circular region of 
0.15$^\circ$ radius around the position of the source.

The obtained spectrum is well described by a power-law model and the integral flux above 1 TeV of 
$\Phi(\rm E>1\rm~TeV)~=~(1.7~\pm 0.9_{\rm stat}~\pm 0.3_{\rm syst})~\times 10^{-13}\rm~cm ^{-2}~s^{-1}$ 
corresponds roughly to 0.7$\%$ of the Crab nebula flux above the same energy.

\section{Multi-wavelength counterparts}

\subsection{X-ray observations}\label{xmm}

Following the indication of the H.E.S.S. source, XMM-Newton was used to observe the gamma-ray emission region in March 
2011.
Only 7 ks of data out of 17 ks are usable for analysis due to proton flare contamination.
The data were processed using the XMM-Newton Science Analysis System (v10.0) and the instrumental background
 (both for image and spectra) was derived from a compilation of blank sky observation \cite{carter07} renormalized in 
the 10-12 keV energy band.

A point-like source with hard, highly absorbed spectrum is lying very close to the best fitted position of the TeV 
emission.
The center of the X-ray source is indeed 30'' away from the fitted position of the H.E.S.S. excess, but both positions are 
compatible within the statistical errors.
No extended emission from non-thermal X-rays was found in these data.
The unabsorbed flux between 2 and 10 keV of the point source is found to be 
$\Phi$(2-10 keV )$= 6^{+3}_{-5} \times 10^{-13}$ erg cm$^{-2}$ s$^{-1}$, with a spectral index of 
$\Gamma=1.3^{+0.6}_{-0.5}$ and a column density of 
N${_H}= 6.9^{+2.8}_{-2.1} \times 10^{22}$cm$^{-2}$. The latter is three to six times larger than the total Galactic 
H~$\textsc{i}$ column density \cite{kalberla} averaged on a cone of $0.5^\circ$ radius around the gamma-ray source 
position.
The quite large derived column density indicates that the source is possibly distant and likely to lie behind dense 
molecular material (one of the clouds presented in the next section).
The point-like source has spectral features which mimic a pulsar-like spectrum but no variability is found in the light 
curve of X-ray data.
However, due to the weak statistics, no strong constraints could be derived.

\subsection{$^ {13}$CO observations: association with molecular clouds}

\begin{figure}[htbp]
\includegraphics[width=70mm]{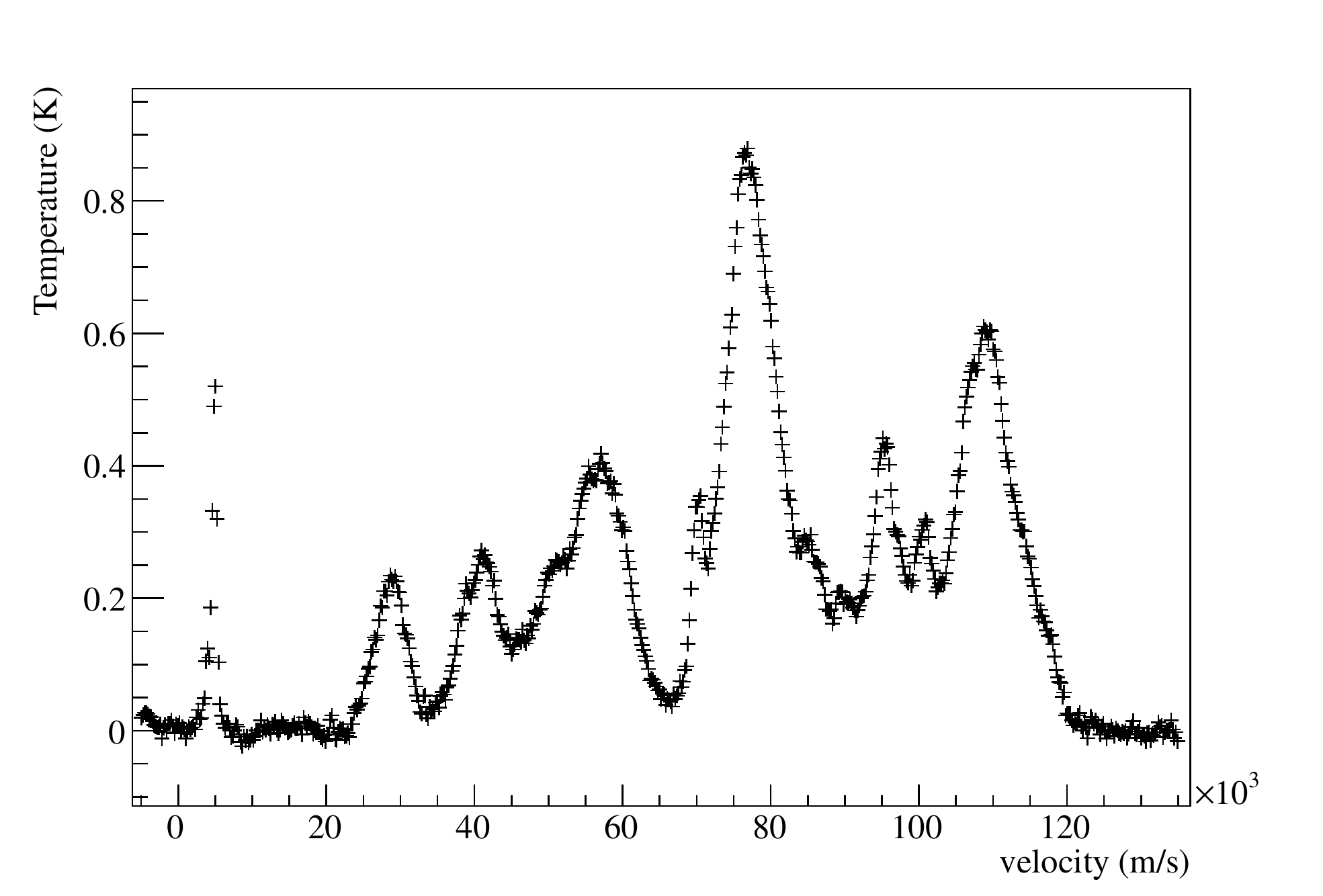}
\caption{Average $^{13}$CO (J=1$\rightarrow$0) transition line observed in the GRS \cite{grs} around the position of the 
TeV emission in a 0.2$^\circ$ side square region.}
\label{13COclouds}
\end{figure}

The Galactic Ring Survey (GRS) performed with the Boston University FCRAO telescopes \cite{grs} provides measurements 
of the $^{13}$CO (J=1$\rightarrow$0) transition line covering the velocity range from -5 to 135 km s$^{-1}$ in this 
region. The detection of this line is evidence for the presence of dense molecular clouds that are known to be 
cosmic-ray targets and hence gamma-ray emitters via neutral pion production and decay.
A square region of 0.2$^\circ$ side is defined around the source HESS J1832-093 to look for spatially coincident molecular 
clouds traced by the $^{13}$CO transition line. Several molecular clouds are found in this region, as it can be seen in 
Fig. \ref{13COclouds} showing the antenna temperature as a function of the measured radial velocity of the 
$^{13}$CO transition line.

\begin{figure*}[htbp]
\begin{minipage}{0.5\linewidth}
\centering
\includegraphics[width=80mm]{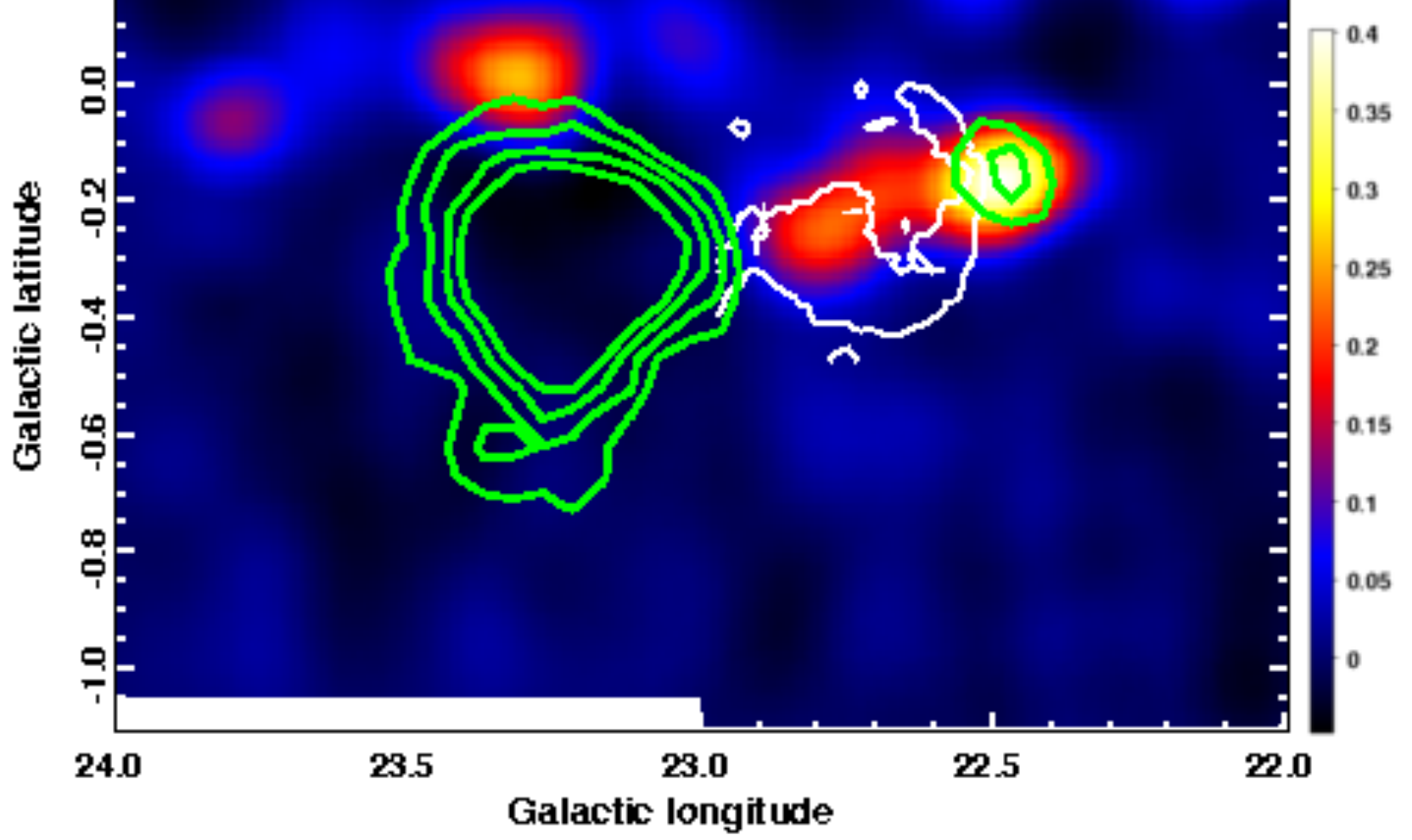}
\end{minipage}%
\begin{minipage}{0.5\linewidth}
\centering
\includegraphics[width=80mm]{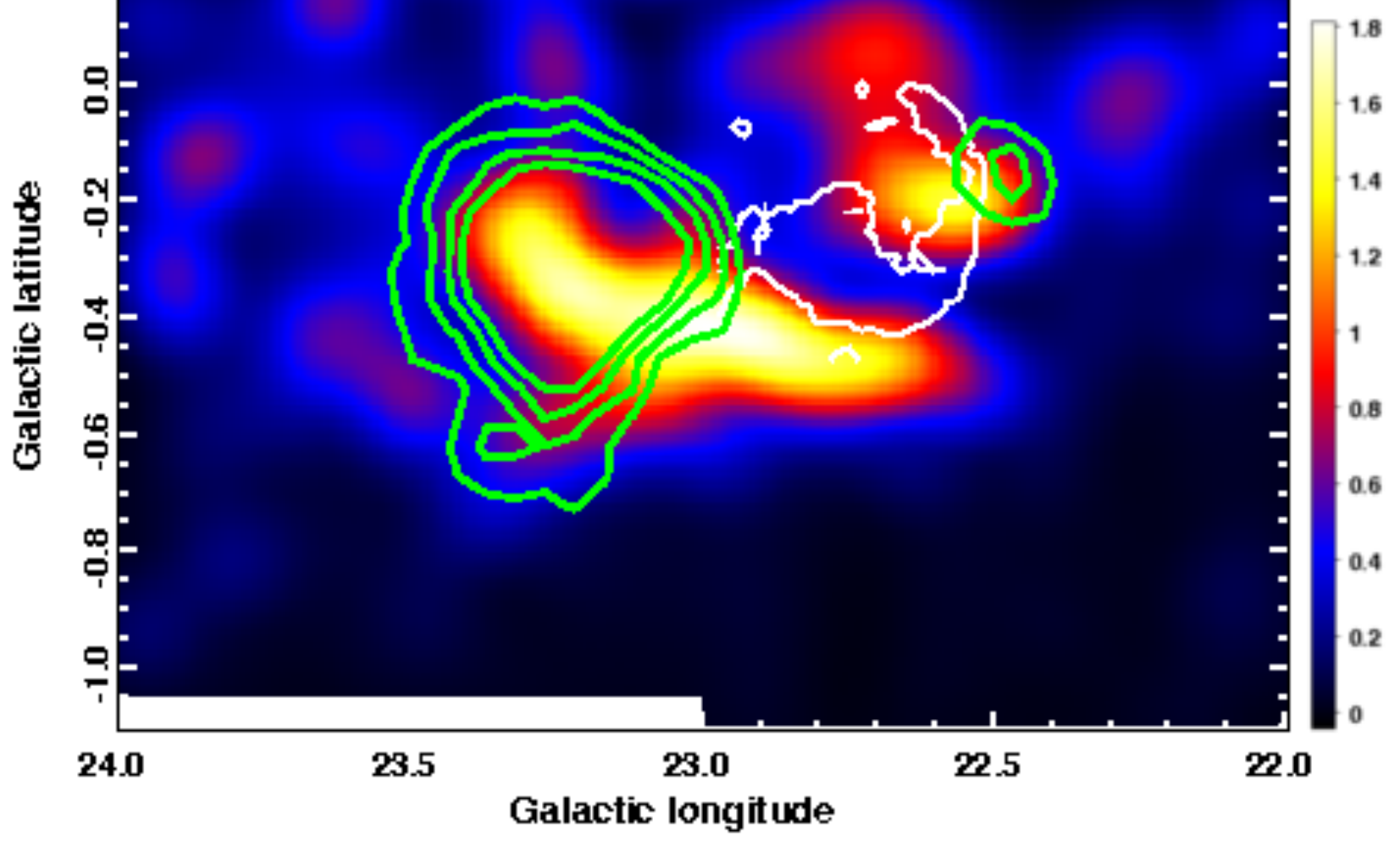}
\end{minipage}
\centering
\caption{$^{13}$CO antenna temperature \cite{grs} in color at 28 km s$^{-1}$ for cloud (a) (left) and 77 km s$^{-1}$ for 
cloud (b) (right) smoothed with the average H.E.S.S. PSF for this data set. The gamma-ray significance contours from 
4$\sigma$ to 7$\sigma$ are shown in thick green while the radio observation of SNR G22.7-0.2 \cite{magpis} is overlaid 
in thin white contours.}
\label{cloud1}
\end{figure*}

Each peak represents a molecular cloud, the mean velocity of a peak is related to the velocity of the considered cloud 
and its width characterizes the velocity dispersion due to the molecules agitation inside the cloud.
The measured radial velocities can then be translated into distances using the Galactic rotation curve model provided by 
\cite{clemens}. Given our position in the galaxy, two distances are therefore possible for each velocity.
The molecular clouds that are most spatially coincident with the gamma-ray source are selected, corresponding to peaks 
(a) and (b) in Fig.~\ref{13COclouds}. These clouds are shown in color in Fig. \ref{cloud1} with overlaid VHE gamma-ray 
significance contours.
We use the associated antenna temperatures to estimate the cloud mass for each possible distance, as described in 
\cite{simon}.
An equivalent radius R$_{eq}$ is then determined from the apparent size of the clouds, with which we define a sphere to 
approximate the volume of each cloud, allowing us to calculate the cloud densities. The obtained parameters for both 
clouds shown in Fig. \ref{cloud1} are listed in Table~$\textsc{i}$.


\vspace{-0.2cm}

\begin{table}[htbp]
\begin{center}
\begin{tabular}{|l|c|c|}
\hline \textbf{Parameters} & \textbf{Close distance} & \textbf{Far distance}
\\
\textbf{Cloud (a)} & 3.2 kpc & 13.8 kpc \\
\hline R$_{eq}$ & 7 pc & 31 pc \\
\hline Mass  & 1.2 $\times$ 10$^3$ M$_{\odot}$ & 2.3 $\times$ 10$^4$ M$_{\odot}$\\
\hline Density & 15 cm$^{-3}$ & 4 cm$^{-3}$ \\
\hline \textbf{Parameters} & \textbf{Close distance} & \textbf{Far distance}
\\
\textbf{Cloud (b)} & 4.8 kpc & 10.5 kpc \\
\hline R$_{eq}$ & 13 pc & 28 pc \\
\hline Mass  & 2.7 $\times$ 10$^4$ M$_{\odot}$ & 1.3 $\times$ 10$^5$ M$_{\odot}$\\
\hline Density & 64 cm$^{-3}$ & 30 cm$^{-3}$ \\
\hline
\end{tabular}
\label{cloudpar}
\caption{: Equivalent radius, mass and density of clouds (a) and (b) for the two possible distances \cite{clemens}.}
\end{center}
\end{table}

\vspace{-1cm}

\section{Discussion}

The position of the gamma-ray emission lies on the edge of the remnant G22.7-0.2 and is spatially coincident with
molecular clouds, suggesting different possible scenarios.
First, cosmic rays that could be accelerated in the remnant might interact with a molecular cloud and produce neutral 
pions, their decay leading to the observed TeV emission. A second scenario involves electrons accelerated in the
 remnant with the production of VHE gamma-rays via Bremsstrahlung radiation in the molecular cloud.

To test the hadronic scenario, we use the parameters derived for the clouds in Table~$\textsc{i}$
and the gamma-ray flux above 1 TeV obtained with the spectral study (see section \ref{hessobs}). 
The expected gamma-ray flux in such a scenario is \cite{dav}:

\begin{eqnarray}\nonumber\label{eq:units}
\textstyle F_{\gamma}(E>E_{0}) &&\approx 9 \times 10^{-11} ~ \theta  ~ \bigl (\frac{E_{0}}{\rm{1~TeV}} \bigr) ^{-1.1}
\bigl(\frac{E_{SN}}{10^{51}\rm{erg}}\bigr)\\
&& \times \bigl(\frac{d}{\rm{1~kpc}}\bigr)^{-2}\bigl(\frac{n}{\rm{1~cm}^{-3}}\bigr) ~ \rm cm^{-2}s^{-1}
\end{eqnarray}

where $\theta$ is the fraction of the total kinetic energy released by the supernova ($E_{SN}$) going into accelerated 
cosmic rays energy, $d$ denotes the distance to the object and $n$ the density of the cloud.
As we know the gamma-ray flux above 1 TeV and derived a density for a given distance, we can compute the $\theta$ fraction 
considering a supernova energy of $10^{51}$ erg. Since the gamma-ray excess is situated only on the edge of the 
remnant, we also included the solid angle of the clouds seen by the SNR. The $\theta$ fractions obtained are respectively 
0.3$\%$ and 0.6$\%$ for clouds (a) and (b) for the closest distance, 27$\%$ and 6$\%$ for the furthest distances. 
Except for the far distance of cloud (a), the derived fractions are well below the 10$\%$ usually taken in the models.
However, cloud (b) seems favoured in this scenario since the computed density and mass are higher. Thus, with more targets 
for the cosmic-rays, it is more likely to be the origin of the VHE gamma-ray emission. Moreover, the obtained distance of 
4.8 kpc is quite close to the distance of W41 that is supposed to be in interaction with G22.7-0.2 as suggested in 
\cite{spitzer}. On the other hand, cloud (a) is more spatially coincident with the gamma-ray emission than cloud (b) 
(see Fig. \ref{cloud1}). 

As mentioned in section \ref{xmm}, a point source with pulsar-like spectrum shape has been found very close to 
the best fitted position of the H.E.S.S. emission. 
The high column density derived could imply a possibly distant source. 
The point source in X-rays is seen neither in IR \cite{spitzer}, nor in radio at 21cm and 90cm wavelengths \cite{magpis}.
 Although the proximity of the X-ray source and the best fitted position of 
the TeV source is striking, the lack of X-ray pulsations does not support a scenario based on pulsar wind nebula 
(PWN). Besides, we did not find any PWN-like emission in the X-ray data that could come from a potential nebula 
associated with the point-source. 
However, this assumption cannot be ruled out since the flux from a distant PWN would be very weak and the 
very short observation time does not allow us to see such a faint emission. Furthermore, giving the high column density 
measured for the point-source, the emission coming from the hypothetical PWN could have been partially absorbed along the 
line of sight. 

No Fermi source is found at the position of the H.E.S.S. excess, but the Fermi source 2FGL J1834.3-0848, potentially 
associated with the remnant W41, is situated less than 1$^\circ$ away from HESS J1832-093. The angular resolution of the 
Fermi LAT could thus prevent to distinguish between the two sources.

\section{Conclusion}

Observations in the field of view of the supernova remnant G22.7-0.2 with the H.E.S.S. telescope array have led to 
evidence for a TeV emission in positional coincidence with a fraction of the SNR radio shell. 
The source HESS J1832-093 reaches a significance level of 5.8$\sigma$ applying standard Hillas-based cuts \cite{crab}. 
It appears slightly extended with an intrisic size of $0.06^\circ \pm 0.02^\circ$. 
The source is quite faint since the integrated flux above 1 TeV is about 0.7$\%$ of the Crab nebula flux above the same 
energy.

The spatial coincidence with $^{13}$CO enhancements measured through the GRS \cite{grs} suggests an interaction with a 
molecular cloud, thus implying hadrons or electrons accelerated in the SNR G22.7-0.2 that would produce gamma-rays in the 
considered cloud by neutral pions production or Bremsstrahlung radiation respectively. 
The latter scenario was not investigated here. 
We have shown that the hadronic scenario is energetically achievable by computing the fraction of the supernova 
kinetic energy required to produce the observed gamma-ray flux using Eq.~\ref{eq:units} \cite{dav}.

On the other hand, recent observations in the X-ray band with the XMM-Newton satellite have brought another possible 
explanation for the observed TeV emission, even though the lack of information does not allow us to deeply investigate 
that scenario.
The discovered point-like source presents a hard spectrum and lies 
within the best fitted position of the gamma-ray excess. This source has pulsar-like spectral features but no 
pulsations were found in the X-ray data.
The large colum density indicates that it should be quite distant, thus, considering the very
 short observation time, we don't expect to see any extended emission coming from a possible nebula associated to the 
X-ray source. More observations to determine the nature of the X-ray source are crucial to further investigate this 
possible scenario.

\bigskip

\bigskip 
\begin{acknowledgments}

The support of the Namibian authorities and of the University of Namibia
in facilitating the construction and operation of H.E.S.S. is gratefully
acknowledged, as is the support by the German Ministry for Education and
Research (BMBF), the Max Planck Society, the French Ministry for Research,
the CNRS-IN2P3 and the Astroparticle Interdisciplinary Programme of the
CNRS, the U.K. Science and Technology Facilities Council (STFC),
the IPNP of the Charles University, the Polish Ministry of Science and 
Higher Education, the South African Department of
Science and Technology and National Research Foundation, and by the
University of Namibia. We appreciate the excellent work of the technical
support staff in Berlin, Durham, Hamburg, Heidelberg, Palaiseau, Paris,
Saclay, and in Namibia in the construction and operation of the
equipment.

This publication makes use of molecular line data from the Boston University-FCRAO Galactic Ring Survey (GRS). 
The GRS is a joint project of Boston University and Five College Radio Astronomy Observatory, funded by the 
National Science Foundation under grants AST-9800334, AST-0098562, \& AST-0100793

\end{acknowledgments}
\vspace{-0.6cm}

\end{document}